\documentclass[journal=jpcafh]{achemso}

\usepackage[version=3]{mhchem} 
\usepackage{fontenc}
\usepackage{chemformula}
\usepackage{amsmath}
\usepackage{amsfonts}
\usepackage{amssymb}
\usepackage{lineno}
\usepackage{mathtools}
\usepackage{xcolor}
\usepackage{setspace}
\usepackage{pdfpages}

\author{F. Mazzola\footnote{These authors contributed equally to this work.} }
\affiliation{CNR-IOM, Area Science Park, Strada Statale 14 km 163.5, I-34149 Trieste, Italy}
\email{mazzola@iom.cnr.it}

\author{S.K. Chaluvadi$^{a}$}
\affiliation{CNR-IOM, Area Science Park, Strada Statale 14 km 163.5, I-34149 Trieste, Italy}

\author{V. Polewczyk}
\affiliation{CNR-IOM, Area Science Park, Strada Statale 14 km 163.5, I-34149 Trieste, Italy}
 
\author{D. Mondal}
\affiliation{CNR-IOM, Area Science Park, Strada Statale 14 km 163.5, I-34149 Trieste, Italy}

\author{J. Fujii}
\affiliation{CNR-IOM, Area Science Park, Strada Statale 14 km 163.5, I-34149 Trieste, Italy}

\author{P. Rajak}
\affiliation{CNR-IOM, Area Science Park, Strada Statale 14 km 163.5, I-34149 Trieste, Italy}

\author{M. Islam}
\affiliation{CNR-IOM, Area Science Park, Strada Statale 14 km 163.5, I-34149 Trieste, Italy}

\author{R. Ciancio}
\affiliation{CNR-IOM, Area Science Park, Strada Statale 14 km 163.5, I-34149 Trieste, Italy}

\author{L. Barba}
\affiliation{Istituto di cristallografia del CNR, Strada Statale 14 km 163.5, I-34149 Trieste, Italy}

\author{M. Fabrizio}
\affiliation{International School for Advanced Studies (SISSA), Via Bonomea 265, I-34149 Trieste, Italy}

\author{G. Rossi}
\affiliation{University of Milano, Via Celoria 16, I-20133 Milano, Italy}
\alsoaffiliation{CNR-IOM, Area Science Park, Strada Statale 14 km 163.5, I-34149 Trieste, Italy}

\author{P. Orgiani}
\affiliation{CNR-IOM, Area Science Park, Strada Statale 14 km 163.5, I-34149 Trieste, Italy}

\author{I. Vobornik}
\affiliation{CNR-IOM, Area Science Park, Strada Statale 14 km 163.5, I-34149 Trieste, Italy}

\title[A genuine Mott transition in V$_{2}$O$_{3}$]
  {Disentangling structural and electronic properties in V$_{2}$O$_{3}$ thin films: a genuine non-symmetry breaking Mott transition}

\abbreviations{ARPES, PLD, XRD. HRTEM}

\begin{document}



\newpage

\begin{abstract}

Phase transitions are key in determining and controlling the quantum properties of correlated materials. Here, by using the powerful combination of precise material synthesis and angle resolved photoelectron spectroscopy, we show evidence for a genuine Mott transition undressed of any symmetry breaking side effects in the thin-films of V$_{2}$O$_{3}$. In particular, and in sharp contrast with the bulk V$_{2}$O$_{3}$ crystals, we unveil the purely electronic dynamics approaching the metal-insulator transition, disentangled from the structural transformation that is prevented by the residual substrate-induced strain. On approaching the transition, the spectral signal evolves surprisingly slowly over a wide temperature range, the Fermi wave-vector does not change, and the critical temperature appears to be much lower than the one reported for the bulk. Our findings are on one side fundamental in demonstrating the universal benchmarks of a genuine non-symmetry breaking Mott transition, extendable to a large array of correlated quantum systems and, on the other, given that the fatal structural breakdown is avoided, they hold promise of exploiting the metal-insulator transition by implementing V$_{2}$O$_{3}$ thin films in devices. 

\end{abstract}


\newpage

The ability of manipulating electronic states in quantum matter is a milestone for the condensed matter physics community. Materials with properties lying at the verge of an instability have attracted attention because even small external stimuli could drive them into completely different electronic and/or magnetic configurations. V$_{2}$O$_{3}$ is a typical example: upon lowering the temperature - in bulk form \cite{Dernier1970} - it exhibits a metal-insulator transition (MIT) accompanied by anti-ferromagnetism and a rhombohedral-monoclinic structural transformation \cite{Leonov2015,McWhan1969,McWhan1973,Paolasini1999,Meneghini2005}. The attention garnered by this material is not fortuitous. Such a MIT spans over ten orders of magnitude in resistivity in a hysteretic fashion, which is fundamental for applications\cite{Yuan2021,Lee2021,Ronchi2019,Zhou2015,Luo2020} in electronics such as oscillators, neuromorphic devices, and memory – yet the impossibility of tuning such a transition due to the structural breakdown has hindered its exploitation in oxide-electronics. Moreover, it was for long considered to be the only physical realisation of a genuine Mott transition, exemplified, e.g., by the single-band Hubbard model \cite{MOTT1968,Dillemans2014}. However, its structural complexity, the multi-orbital nature and the concurrent metal-insulator, paramagnetic-antiferromagnetic and rhombohedral-monoclinic transitions, have challenged that simple picture. Indeed, a combined LDA+DMFT calculation by Poteryaev et al. \cite{Poteryaev2007} unveiled a mechanism for the paramagnetic MIT in V$_{2}$O$_{3}$, which was mostly driven by the orbital degrees of freedom. Specifically, electronic correlations were shown to substantially enhance the low-energy effective trigonal crystal-field splitting between the lower e$^{\pi}_{g}$ doublet and the upper a$^{1}_{g}$ singlet. Such an enhancement leads to a nearly empty a$^{1}_{g}$ electron pocket at the Fermi level, and a highly incoherent nearly half-filled e$^{\pi}_{g}$ band amenable to Mott’s localization and magnetism. This mechanism, which seems to explain observed photo-induced insulator-to-metal transitions \cite{Ronchi2019,Lantz2017}, has been later questioned by angle-resolved photoemission spectroscopy (ARPES) data in metallic V$_{2}$O$_{3}$ at 200 K, i.e., above the T$_{MIT}$ of 165\,K \cite{LoVecchio2016}. Here, a Fermi surface composed by both a$^{1}_{g}$ and e$^{\pi}_{g}$ was found, not compatible with a system at the verge of an a$^{1}_{g}$‑e$^{\pi}_{g}$ gap opening and selective e$^{\pi}_{g}$ Mott’s localization. Therefore, despite considerable efforts, the mechanism leading to the MIT of V$_{2}$O$_{3}$ and its claimed entanglement with the rhombohedral-to-monoclinic structural transitions remains an unsolved mystery after more than century of extensive investigation.

Here, by using strain-engineering thin film technology, we froze the crystal structure of thin V$_{2}$O$_{3}$ films and thus succeeded in studying the pure electronic behaviour of the system undergoing the MIT disentangling it from the structural transition. After performing precise thin-film growth via in-situ pulsed laser deposition (PLD), we exploited ARPES with polarized synchrotron-light to identify the dominant orbital character of the measured electronic bands. Finally, we disclosed the full temperature evolution of the spectroscopic features approaching the MIT, providing a strong experimental evidence for a genuine Mott transition, void of symmetry breaking. Transport measurements and temperature dependent X-ray diffraction (XRD) confirm a hysteretic behavior in the resistivity with the lack of any structural changes within the same temperature interval. This indicates that the onset of the hysteresis has a purely electronic nature, in agreement with previous thin-films transport measurements \cite{Majid2017} but opposite from what conjectured for the bulk \cite{McLeod2017}. In our ARPES data, the transition critical point is preceded by a continuous and gradual disappearance of spectral weight at the Fermi level, accompanied by a lack of k$_{F}$ variation, consistent with avoided structural transition. This trend on approaching the critical temperature is consistent with the Mott transition described by DMFT \cite{Georges1996}, and in contrast to the abrupt first order character observed for bulk V$_{2}$O$_{3}$. 

V$_{2}$O$_{3}$ films were grown on (0001)-oriented Al$_{2}$O$_{3}$ by PLD \cite{Chaluvadi2021}. The sample temperature was kept at $\sim 700^{\circ}$C and in an oxygen background pressure of $7 \cdot 10 ^{-7}$\,mbar throughout the growth. After deposition, films were cooled down to room temperature under the same deposition pressure. The typical deposition rate was about 3.5\,\AA $\cdot$min$^{-1}$ thus allowing a full control of the film thickness. All of the investigated samples were 15\,nm thick. In-situ X-ray Absorption Spectroscopy (XAS) did not show any line-shape difference in the V L$_{2,3}$ edges from the reference bulk material and the V$_{2}$O$_{3}$ films at room temperature \cite{Caputo2022}.

\begin{figure}[!t]
\centering
\includegraphics[width= \columnwidth]{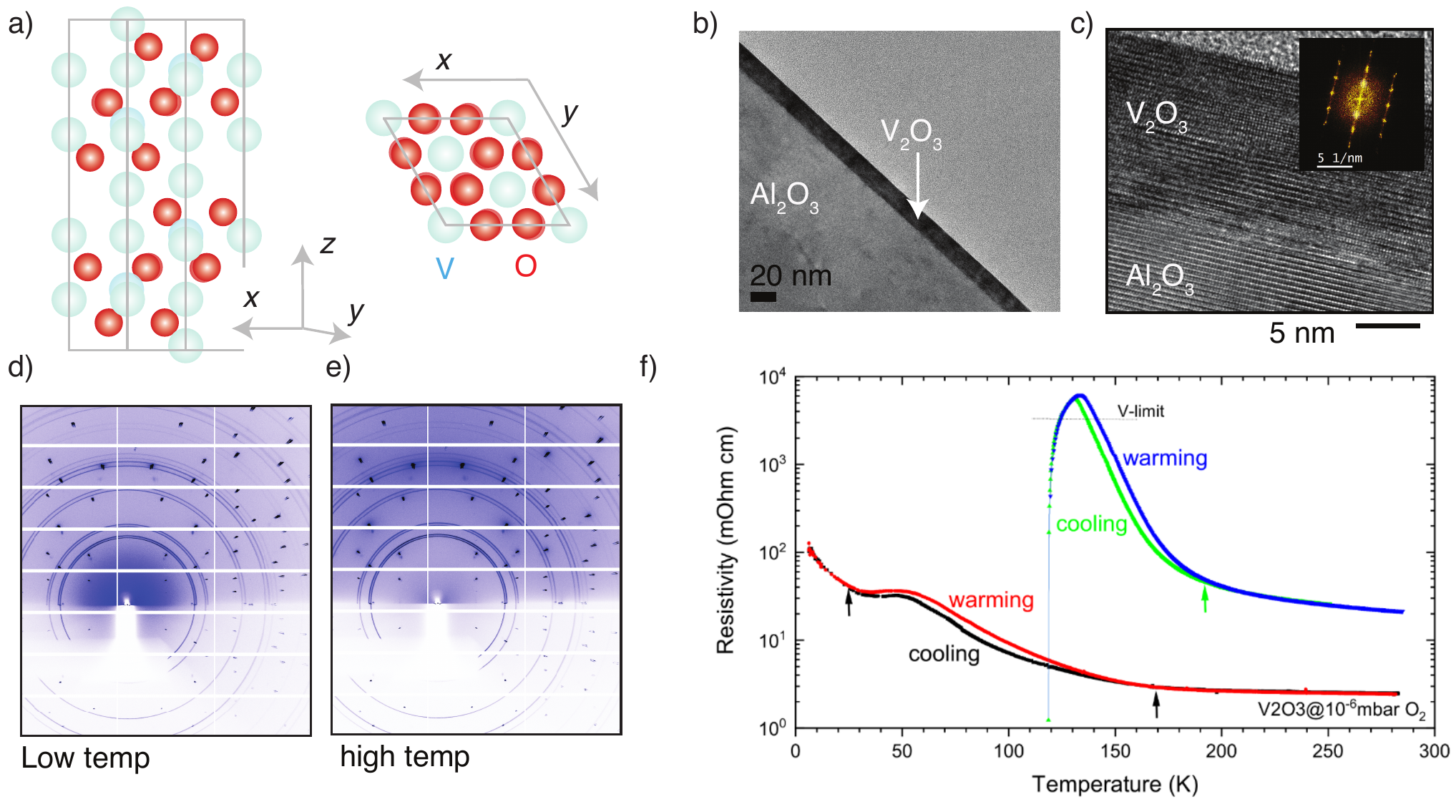}
\caption{\small{(a) Crystal structure of V$_{2}$O$_{3}$ along with the real space unit cell vectors. Both side and top view are reported. (b-c) Low- and High-magnification HRTEM micrographs of V$_{2}$O$_{3}$ on Al$_{2}$O$_{3}$ (in inset the Fourier Transform of the image), respectively. (d-e) Grazing-Incidence XRD maps at high- (i.e. room temperature) and low-temperature (i.e. about 96\,K on the sample) of a representative 15 nm V$_{2}$O$_{3}$ film on Al$_{2}$O$_{3}$ (diffraction rings are related to residual silver paste used to ground the sample and on the back of substrate).}}
\end{figure}

 Structural properties of V$_{2}$O$_{3}$ films were investigated by ex-situ XRD. While thick films (e.g. 80\,nm)\cite{Caputo2022,Giorgianni2022} behave like bulk samples \cite{Rozier2002}, in thinnest (i.e. <\,45\,nm) films, the out-of-plane lattice parameter shifted to lower values (i.e. 13.92\AA with respect to the bulk value of 14.0161\AA). The restoring of the bulk-like properties of V$_{2}$O$_{3}$ in very thick films, clearly indicates the crucial role of the substrate in determining the structural properties of the thin ones. The atomic structure of the V$_{2}$O$_{3}$ films was investigated by high-resolution transmission electron microscopy (HRTEM). The structure of the film over the whole image is homogeneous, with a very smooth surface and free of significant defects. No structural differences were detected among the near-interface region and far from it, as well as no traces of spurious phases or segregation of crystalline phases other than V$_{2}$O$_{3}$, while structural dislocations are mostly present at film/substrate interface. The fast Fourier transformation (FFT) patterns can be safely assigned to the corundum-phase structure \cite{Rozier2002}. In order to investigate the occurrence of corundum-to-monoclinic phase transition in thin films, variable-temperature Grazing-Incidence XRD (GIXRD) measurements (panels c-d in Fig.1) were performed at the X-ray Diffraction beamline 5.2 at Elettra (Trieste, Italy). GIXRD measurements (see also supplementary information) confirmed the absence of any structural change from high temperature (i.e. 300\,K) down to our minimum temperature allowed by the setup (96\,K on the sample). As a matter of fact, the diffraction pattern remains exactly the same, while major differences would have been observed for a monoclic structural phase at low T \cite{Dernier1970}. This is the crucial difference between our thin films and the bulk sister compounds, where a strong crystal symmetry breaking occurs at 165\,K \cite{Dernier1970}. 

The ARPES measurements were performed in ultrahigh vacuum ($<$1 x 10$^{-10}$\,mbar) at the APE-LE beamline at Elettra, with a Scienta DA30 hemispherical electron energy analyser and with linearly polarized photons of \,72 eV. In our setup, the light impinges on the (001) surface with a 45$^{\circ}$ incidence angle, so that the plane identified by the light wave-vector \textbf{q} and the c-axis of the sample corresponds to a mirror plane of the R¯3c space group of the corundum structure, which we conventionally take as the y-z plane (panel a in Fig.2). The light polarization is either parallel to the axis perpendicular to that mirror plane, in our convention parallel to the x-axis (polarization E$_{s}$ = (1,0,0)), or perpendicular to it and to \textbf{q} (polarization E$_{p}$ = (0,1,1)/$\sqrt{2}$).

\begin{figure}[!t]
\centering
\includegraphics[width= 0.7 \columnwidth]{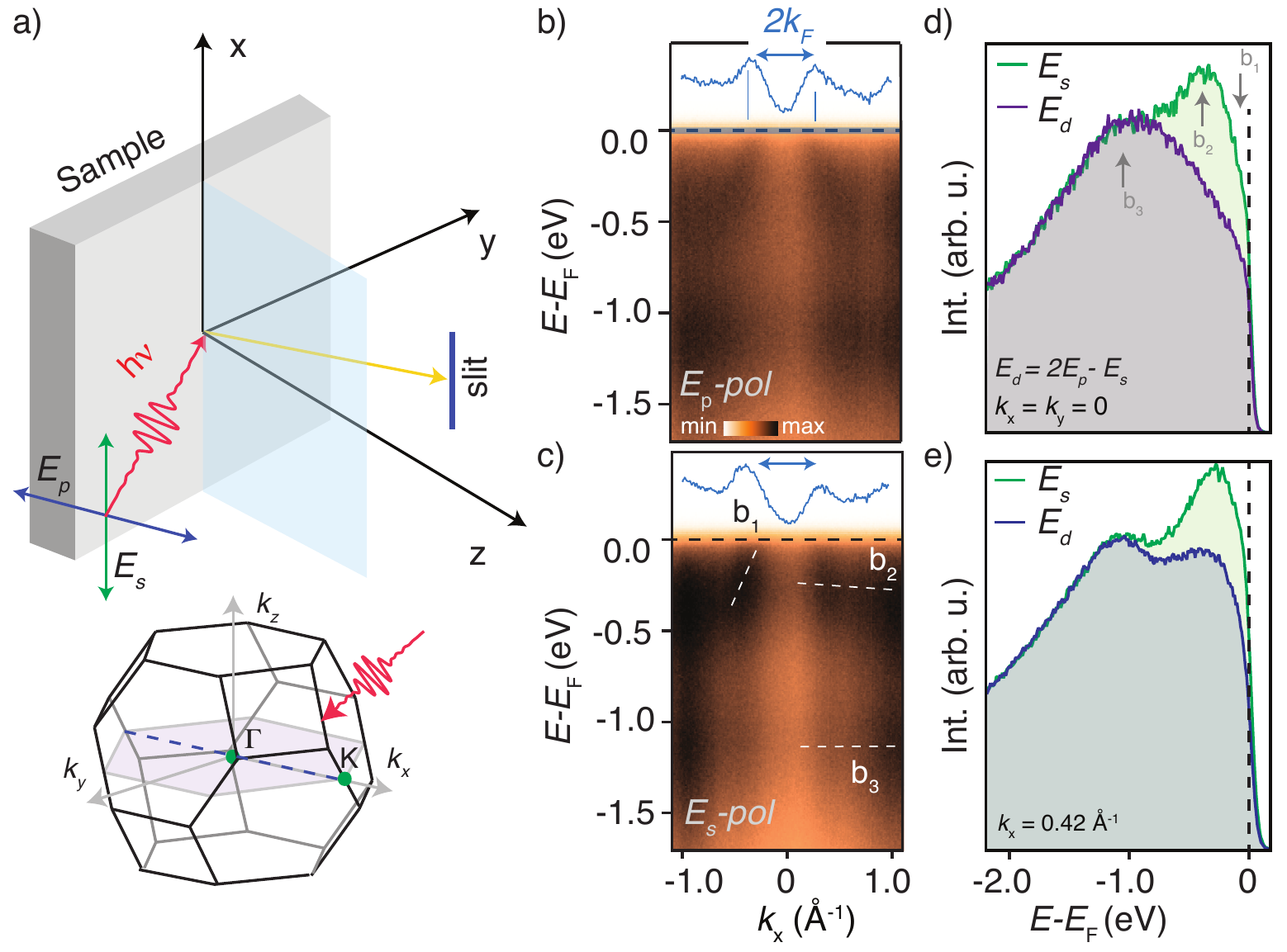}
\caption{\small{(a) Experimental geometry with the relevant scattering plane and light polarization vector and the three-dimensional Brillouin zone. $\Gamma$-K direction is by the green dots. (b-c) ARPES measurements at 230\,K for E$_{s}$- and E$_{p}$-polarizations (h$\nu$ = 72\,eV) showing sensitivity to the orbital character. Three bands are visible and indicated with b$_{1}$, b$_{2}$ and b$_{3}$. (d) $\Gamma$ point (k$_{x}$\,=\,k$_{y}$\,=\,0) E$_{s}$- and E$_{d}$ (2E$_{p}$ - E$_{s}$)-polarizations energy distribution curves showing a strong dichroism for b$_{2}$ which indicates a prominent e$^{\pi}_{g}$ character and a mixed orbital signal (e$^{\pi}_{g}$ and a$^{1}_{g}$) for b$_{3}$. (e) Dichroism spectra for energy distribution curves at k$_{x}$ $\geq$ k$_{F}$ (k$_{y}$\,=\,0, k$_{x}$\,=\,0.42\,\AA$^{-1}$), highlighting the orbital character for b$_{1}$ away from the $\Gamma$ point. The evident difference in the signals allows us to demonstrate the e$^{\pi}_{g}$ character for this band. The same result can be found for any other value of k$_{x}$\,$\geq$\,k$_{F}$.}}
\end{figure}

In our adopted geometry (Fig.2a), the $\Gamma$-K high symmetry direction of the paramagnetic metallic phase of V$_{2}$O$_{3}$ is along  x axis. In this configuration we find that the photoemission matrix elements are favourable and the spectral intensity prominent. The O$_{2p}$ states and the V$_{3d}$ bands are easily identified in the spectra. The formers disperse at high binding energies, with their maximum at 4\,eV below the Fermi energy while the V$_{3d}$ bands extend closer to E$_{F}$ with a bandwidth of about 2\,eV. In the metallic paramagnetic phase, at least three sets of bands are recognizable within the energy range between -1.5\,eV and the Fermi level (Fig.2). We refer to each manifold with b$_{i}$ ($i=1,2,3$), and start discussing their main features at high temperature (i.e. T\,=\,230\,K). In order to better identify the orbital character of the signal, we analyze the polarization dependence at the high symmetry $\Gamma$ point, and from them we can extract the purely in-plane (E$_{s}$) and out-of-plane (E$_{d}$\,=\,2 E$_{s}$\,-\,E$_{p}$) polarization contributions that derive from the e$^{\pi}_{g}$ and a$^{1}_{g}$ orbitals, respectively.

The nearly dispersionless feature b$_{3}$ lies around -1.2\,eV below the Fermi level and has a rather broad photoemission signal which is basically the same for E$_{s}$ and E$_{d}$ light polarizations (see Fig. 2d for the spectra at k$_{x}\,=\,$k$_{y}$\,=\,0, i.e., at the $\Gamma$ point). We associate b$_{3}$ to the lower Hubbard band with all t$_{2g}$ orbitals equally populated, thus an e$^{\pi}_{g}$:a$^{1}_{g}$ occupation ratio of 2:1 compatible with previous results in the low temperature antiferromagnetic insulating phase\cite{Park2000}. The manifold b$_{2}$ lies around -0.3\,eV below the Fermi level and shows a very weak but still evident dispersion. The dependence of its photoemission signal upon light polarization suggests that b$_{2}$ has dominant e$^{\pi}_{g}$ character (Fig. 2d). We do not find evidence that b$_{2}$ crosses the Fermi level upon increasing k$_{z}$ from $\Gamma$ towards the Z point, which would thus lead to the electron pocket observed by Lo Vecchio et al. \cite{LoVecchio2016}, at the (100) surface. We believe that this might be due to the presence of a dead layer \cite{Borghi2009,Rodolakis2009} more pronounced at the (001) surface than at the (100) one (as conjectured in Ref\cite{LoVecchio2016}). As a matter of fact, the dead layer mechanism is more effective for the out-of-plane orbital components, i.e., the a$^{1}_{g}$ and, therefore, it is expected to play a major role at the (001) surface compared to the (100). This is compatible with our evidence that b$_{2}$ has mostly e$^{\pi}_{g}$ character rather than the a$^{1}_{g}$ ones. Finally, the metallic band b$_{1}$ disperses crossing the Fermi level with a nearly circular hole-like Fermi surface of radius 0.36\,$\pm$\,0.02\,\AA$^{-1}$, corresponding to an electron filling fraction of 0.76\,$\pm$\,0.6. Although the maximum of b$_{1}$ is not visible in our data, the polarization dependence of the signal above k$_{F}$ still suggests a prevailing e$^{\pi}_{g}$ character (Fig. 2e). Overall, our high temperature data are in agreement with those reported in literature \cite{LoVecchio2016}. 

\begin{figure}[!t]
\centering
\includegraphics[width= 0.7 \columnwidth]{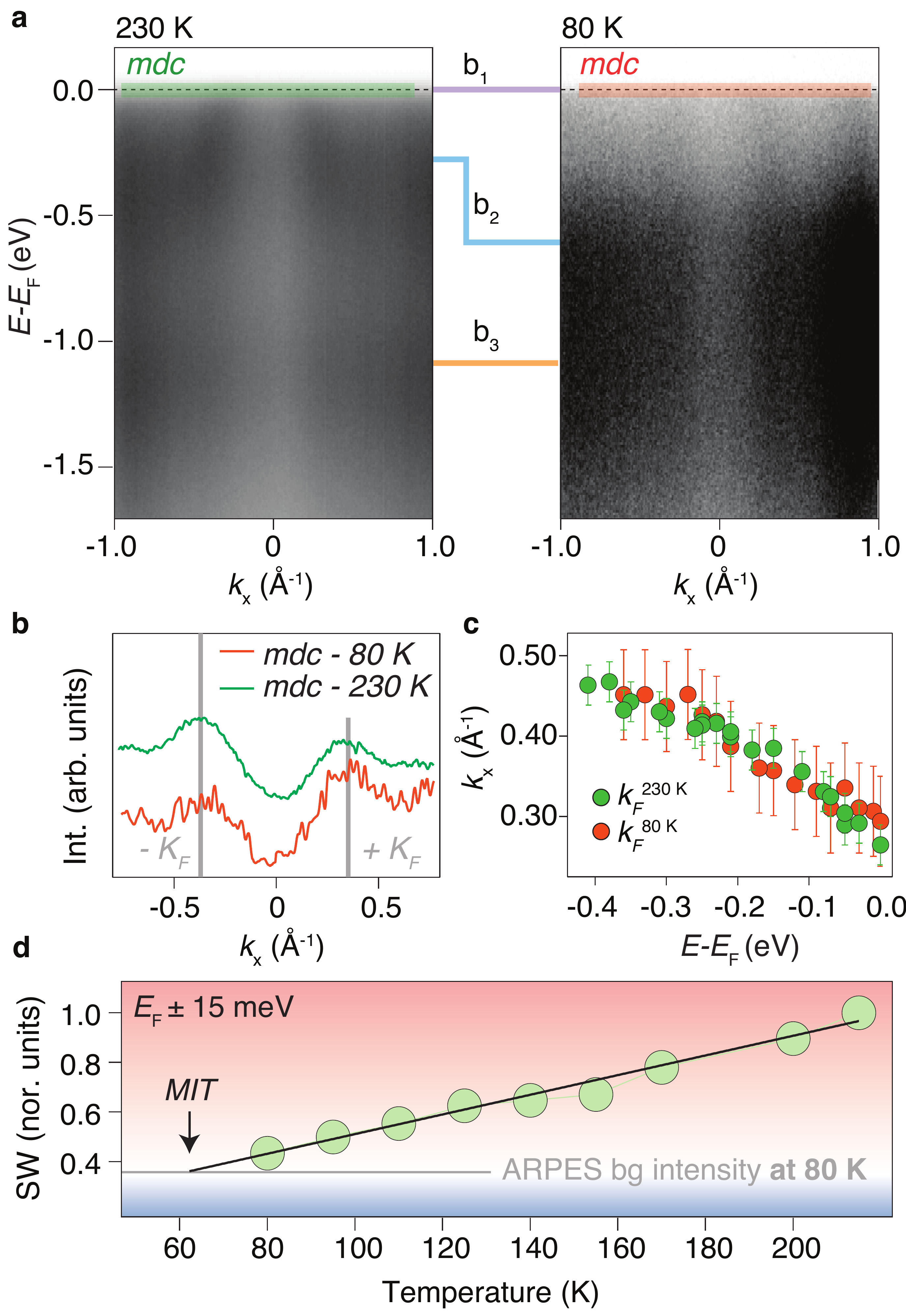}
\caption{\small{(a) ARPES measurements - at 72\,eV - at 230\,K (left) and 80\,K (right). (b) Fermi level wave-vectors reported for both temperatures showing in both cases the same value. This indicates that negligible change in the band b$_{1}$ is detected at the Fermi level. (c) Fitted positions of the band b$_{1}$ (average of left and right branch reported) showing that the band remains the same throughout the temperature range. (d) Trend of the ARPES spectral weight at the Fermi level showing a perfectly linear decrease in the intensity of b$_{1}$. From this trend, we extrapolate that b$_{1}$ will vanish at the extrapolated temperature of 62\,K.}}
\end{figure}

At lower temperatures, significant differences in the measured thin film V$_{2}$O$_{3}$electronic structure arise. ARPES measurements at 80\,K (Fig.3a) mainly show a loss of spectral weight for the b$_{1}$ band and a shift in energy of about 340\,meV of the b$_{2}$ band. Despite the spectral weight loss, we find that b$_{1}$ always crosses the Fermi level at a wave-vector that remains stable from 230\,K down to 80\,K (see Fig. 3b). The spectral weight of b$_{1}$diminishes linearly as a function of temperature, allowing us to extrapolate a tentative metal-insulator transition critical temperature as the one at which the b$_{1}$ spectral weight at E$_{F}$ becomes comparable to the background intensity. Such temperature, as indicated in Fig.3d, is around T$_{MIT}$\,=\,62\,K, which is significantly lower than the bulk value of 165\,K. The decrease of the b$_{1}$ intensity is compensated by a similar increase in the signal of b$_{3}$, as shown in Fig.4 (panel c). However, the b$_{2}$ signal remains constant at all temperatures. The observed transfer of spectral weight is reminiscent of an electronically driven Mott transition. 

\begin{figure}[!t]
\centering
\includegraphics[width= 0.7 \columnwidth]{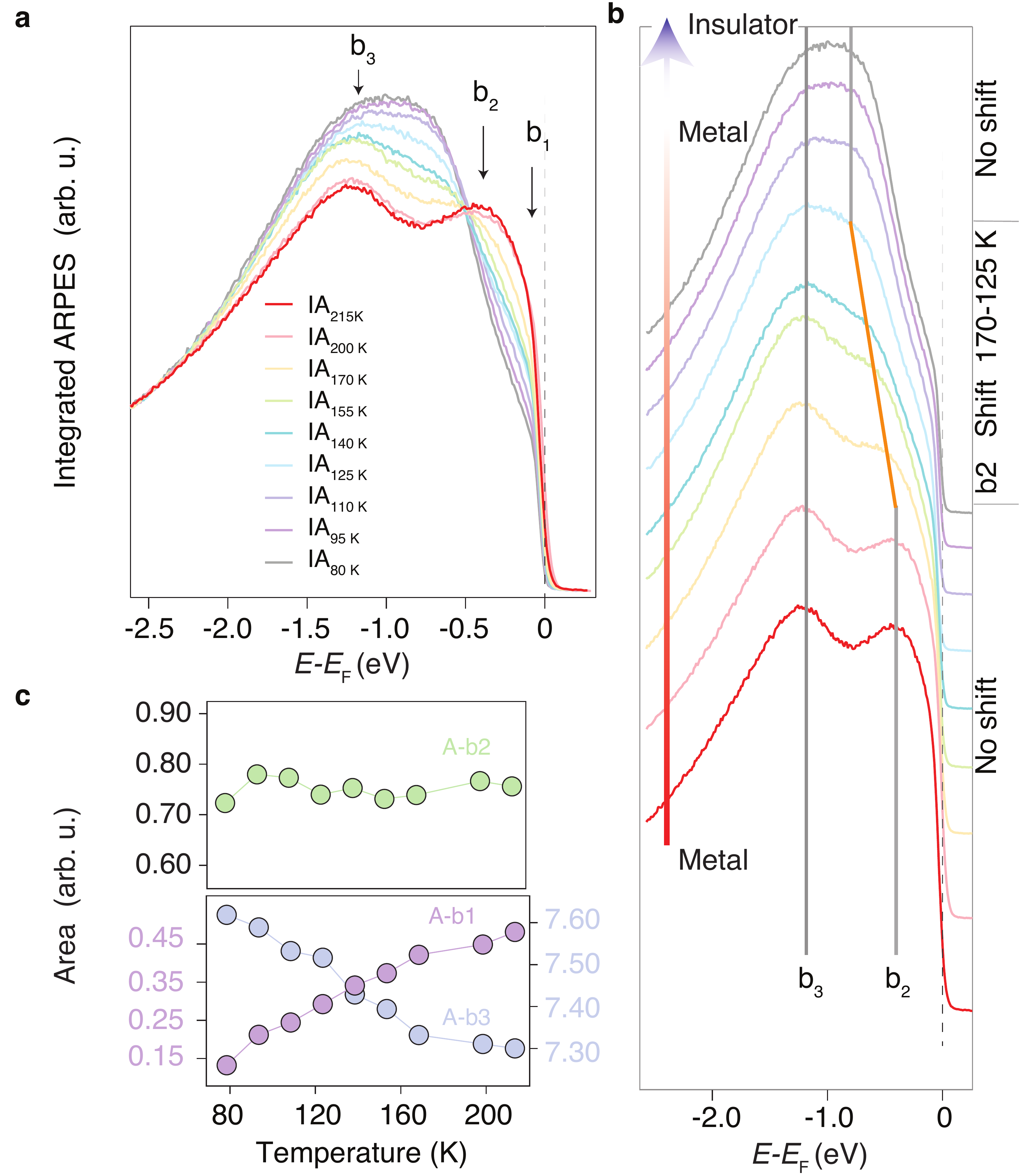}
\caption{\small{(a) Angle-integrated photoemission intensity as function of temperature showing the evolution of the bands detected as function of temperature. (b) Waterfall plot of the integrated intensity showing that the energy shift of the band b$_{2}$ as a function on temperature. (c) Spectral intensity of b$_{1}$, b$_{2}$ and b$_{3}$ as a function on temperatures.}}
\end{figure}

With decreasing temperature, b$_{1}$, b$_{2}$ and b$_{3}$ bands evolve differently while temperature dependent GIXRD measurement show no evidence of the rhombohedral-monoclinic structural transformation reported in bulk and thicker films. Angle-integrated photoemission intensity of the b$_{i}$-bands shows a loss of spectral weight of the band b$_{1}$ and its transfer to the b$_{3}$ lower state (panels a-c in Fig.4). The b$_{3}$ band remains constant in energy and negligible changes are observed in its dispersion, as also highlighted by the constant high-energy tail (Fig. 3a-b). On the contrary, for b$_{2}$ we observe a large downward energy shift. Such a shift occurs between 170\,K and 125\,K and stabilizes below 125\,K. It is possible that this energy shift can still develop further but we do not observe this within our energy and momentum resolutions ($\sim$15\,meV and 0.02\,\AA$^{-1}$, respectively) and the evident broadening of our data. This behavior is suggestive of an effective increase of the trigonal crystal field splitting that pushes down in energy the e$^{\pi}_{g}$ orbital, in agreement with the LDA-DMFT prediction \cite{Poteryaev2007}, although we cannot see a corresponding upward shift of the a$^{1}_{g}$, which is prevented by the dead layer mechanism. What we find remarkable is that such purported enhancement of the trigonal field is not gradual but occurs in a rather narrow temperature range. Even more remarkable is the lack of evidence of a downward shift of the quasiparticle b$_{1}$ band at the Fermi level, which we find has also e$^{\pi}_{g}$ character. This suggests that the transition from the high temperature metal to the low temperature Mott insulator that we observe is not entirely similar to the one predicted from the paramagnetic metal to the paramagnetic insulator upon rising the Hubbard U \cite{Poteryaev2007}. In other words, although the energy shift of the b$_{2}$ band is compatible with an enhancement of trigonal field, such an enhancement seems ineffective at low energy, at least for the e$^{\pi}_{g}$ orbital, possibly because it is compensated by the diminishing quasiparticle residue.

This behavior is summarized in Fig. 5, where, after fitting the k-integrated ARPES intensity, i.e., mimicking the density of states – DOS, with a minimal set of three Lorentzian components (Fig.5a) in Fig. 5b we include the results of the peak energy positions as function of temperature, for both E$_{s}$ and E$_{p}$ light polarisations (the same fitting procedure was also used to obtain the data in Fig.4c).

\begin{figure}[!t]
\centering
\includegraphics[width= 0.75 \columnwidth]{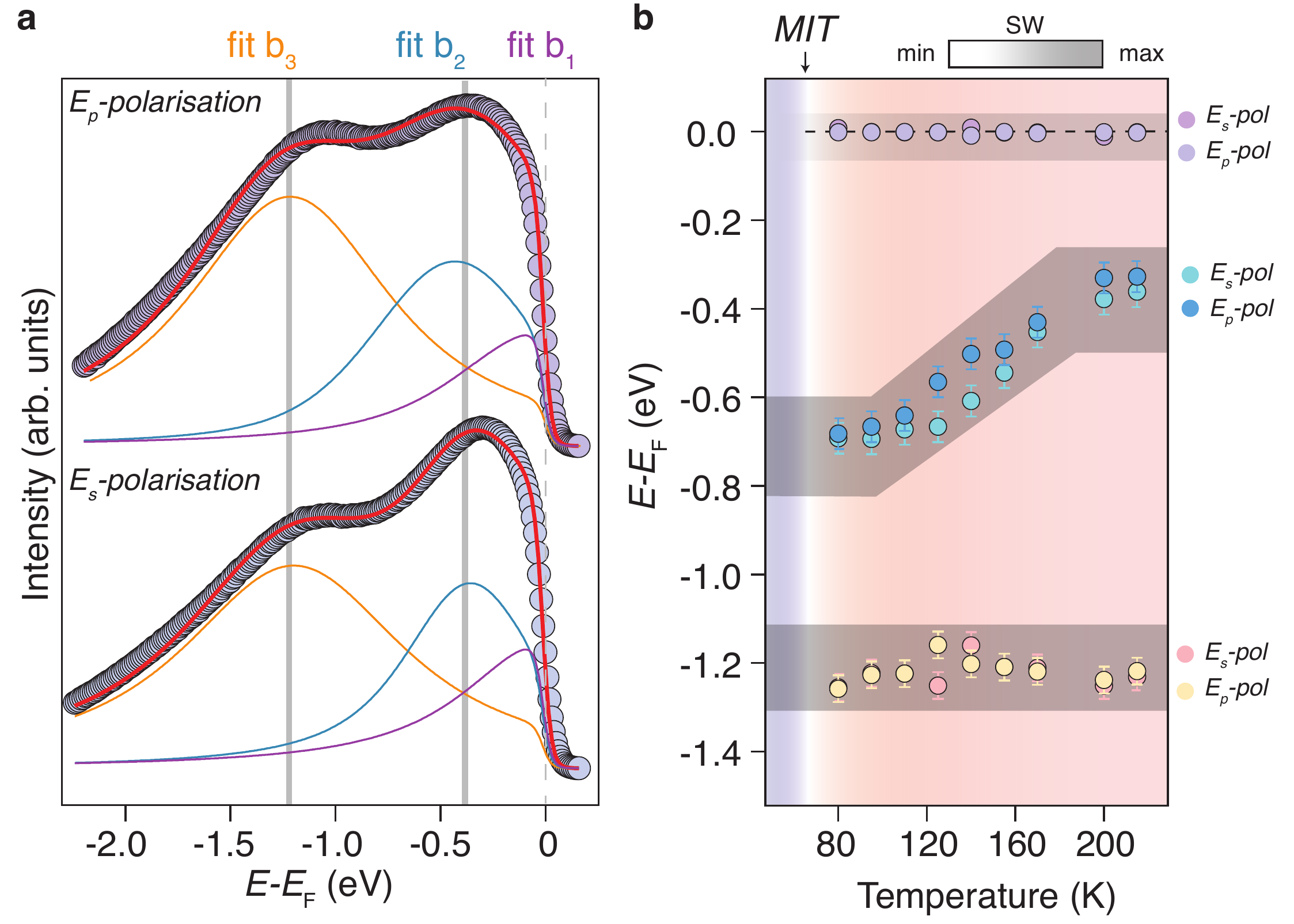}
\caption{\small{(a) Details of the fits executed for the k-integrated ARPES intensity, for both E$_{s}$ - and E$_{p}$-polarised light. The minimum model includes three broad peaks. Each peak is described by a Voigt shape (Lorentzian convoluted by a Gaussian to account for the instrumental resolutions). A Fermi edge has been taken into account. (b) Energy positions of the centroid of the bands b$_{1}$, b$_{2}$ and b$_{3}$.}}
\end{figure}

Should the centre of gravity of b$_{2}$ follow the same linear decrease of b$_{1}$ spectral weight, one could possibly attribute such behaviour to coexisting insulator and metal domains, the fraction of the latter linearly vanishing at the transition. However, the substantial energy shift of b$_{2}$ that occurs in a narrow temperature window relatively to the slow linear decrease of b$_{1}$ spectral weight rules out that scenario \cite{McLeod2017,Thees2021}. In addition, the coexistence of two different electronic environments be present would be directly detected by ARPES \cite{Mazzola2018}. Our observation is therefore reminiscent of a genuine Mott-like transition \cite{Meyer2016,Meyer2016,Brouet2015} that occurs when the band b$_{1}$ that is crossing the Fermi level loses all its spectral weight. We emphasise that such a conclusion has the caveat that, as earlier mentioned, we cannot access the evolution at the Fermi level of the a$^{1}_{g}$ orbital. Importantly, we also found analogous results for samples of similar thickness but grown in a different partial O$_{2}$ pressure with the only effect to move up (when grown at lower O$_{2}$ pressure) and down (when grown at higher O$_{2}$ pressure) the critical temperature at which the b$_{1}$-to-b$_{3}$ transfer of spectral weight occurs \cite{Leiner2019,Thorsteinsson2018}.  

In conclusion, by exploiting in-situ high-precision growth we were able to freeze our V$_{2}$O$_{3}$ this films in the corundum phase and thus avoid the structural transformation occurring in bulk system. This in turn allowed us to observe the purely electronic dynamics across the metal-insulator transition that resembles the textbook example of a non-symmetry breaking Mott transition as revealed by DMFT \cite{Georges1996}. Indeed, the particular low-temperature magnetic order in bulk V$_{2}$O$_{3}$ is believed to be consequence of a substantial magnetic frustration that is resolved only by the C3 symmetry breaking at the rhombohedral-monoclinic transition \cite{Grieger2012,Grieger2015,Leiner2019}. We cannot exclude that, once the structural transformation is circumvented (as in our thin film regime), the magnetic transition is pushed below the metal-insulator one, which would render the observed MIT a genuine paramagnetic Mott transition. Understanding and controlling such an electronic transition is fundamental to enable novel emergent phases of matter, with the confluence of magnetism, correlations and magnetic frustration.


\newpage

\begin{acknowledgement}

FM acknowledges Prof. Sergio di Matteo and Dr. Giancarlo Panaccione for the useful discussions on the topics. Ezio Cociancich, Federico Salvador and Giuseppe Chita are acknowledged for technical support. This work has been performed in the framework of the nanoscience foundry and fine-analysis (NFFA-MIUR Italy Progetti Internazionali) facility. MF has received funding from the European Research Council (ERC) under the European Union's Horizon 2020 research and innovation programme, Grant agreement No. 692670 “FIRSTORM”.

\end{acknowledgement}

\section*{Authors contributions}

F. M., S. K. C. and D. M. performed ARPES experiments and analyzed the ARPES data, with contributions and guidance by G. R., J. F. and I. V.; F. M., S. K. C. and P. O. grew the samples by PLD; L. B. and P. O. measured XRD; P. M., R. I. and R. C. measured and analyzed the TEM data; V. P. performed resistivity measurements. M.F. contributed in theoretical understanding; F.M., M.F., P. O., G. R., J. F. and I. V. wrote the manuscript with contributions from all the authors.

\section*{Data Availability}
Authors can confirm that all relevant data are included in the paper and/ or its supplementary information files.

\section*{Competing Interests}
The authors declare that there are no competing interests.

\newpage

\appendix
\section{Appendix A}



\section*{Supplementary material}

# Appendix A

# Disentangling structural and electronic properties in $V_2O_3$ thin films: a genuine non-symmetry breaking Mott transition

F. Mazzola[1]*[‡], S. K. Chaluvadi[1][‡], V. Polewczyk[1], D. Mondal[1], J. Fujii[1], P. Rajak[1], M. Islam[1], R. Ciancio[1], L. Barba[4], M. Fabrizio[2], G. Rossi[1-3], P. Orgiani[1], I. Vobornik[1]

Supplementary Data

Fig.S1  Temperature dependent GIXRD measurements

Fig.S2  Resistivity data for $V_2O_3$ as a function of thickness and oxygen content

Fig.S3  Overview of the bands that form the electronic structure of $V_2O_3$

Fig.S4  ARPES measurements along the $\Gamma$-K direction at different $k_z$

Fig.S5  ARPES Fermi surface map collected at the plane $k_z = 0$

[1] CNR-IOM, Area Science Park-Basovizza, 34149 Trieste (TS), Italy
[2] International School for Advanced Studies (SISSA), Via Bonomea, 265, 34136 (TS), Italy
[3] Dipartimento di Fisica, Università degli studi di Milano, Via Celoria, 16 20133 MILANO (MI), Italy
[4] CNR-IC, Area Science Park-Basovizza, 34149 Trieste (TS), Italy
[‡] *These authors contributed equally to this work.*

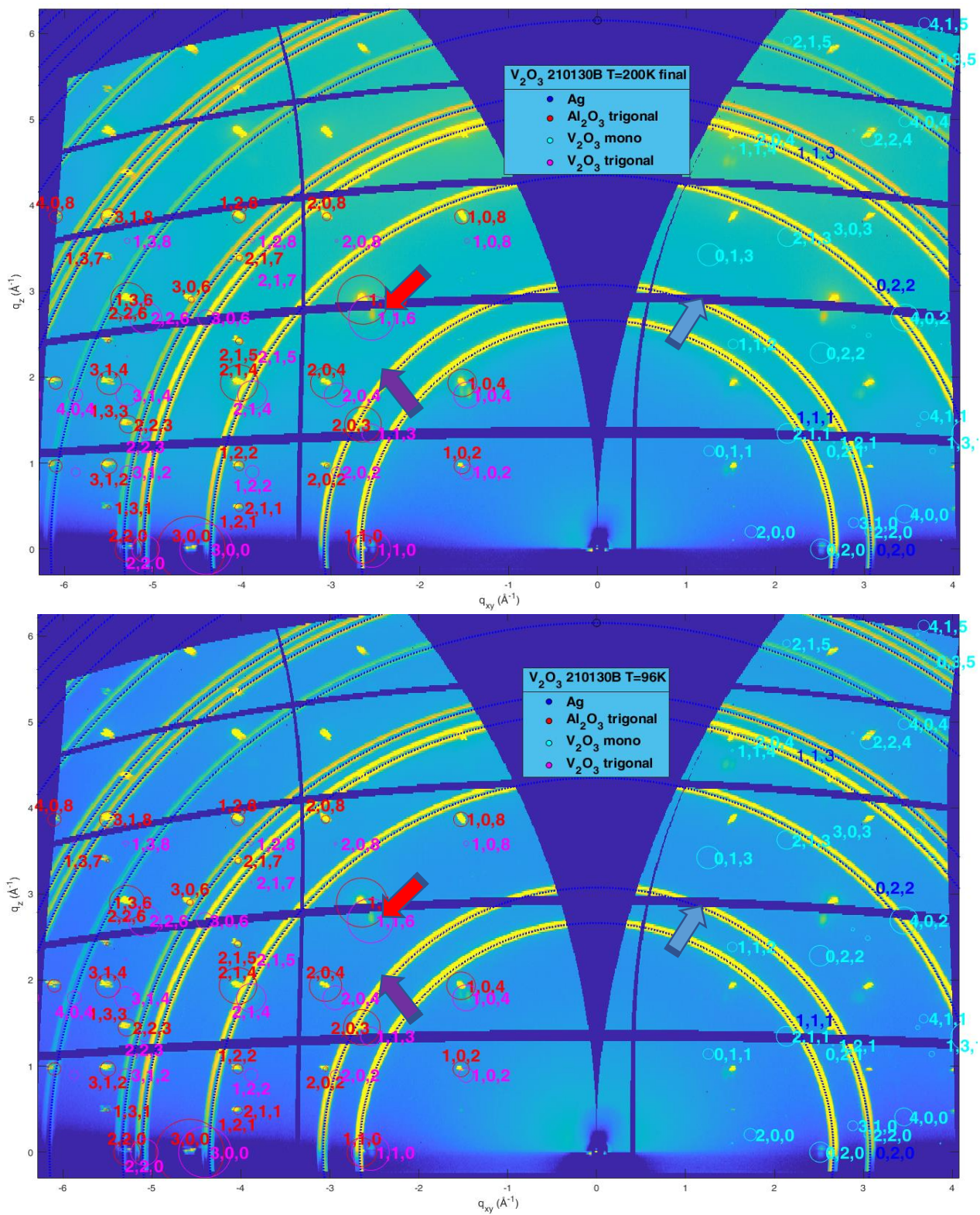

Figure S1. Temperature dependent GIXRD measurements of a V2O3 film at high (i.e. 200K) and low (i.. 96K) temperatures; expected position of diffraction peaks corresponding to trigonal-Al2O3 (red), trigonal V2O3 (magenta) and monoclinic V2O3 (blue); a few selected peaks for the three phases are enlightened by arrows (colour-code is the same). Yellow rings are related to residual silver paste underneath the substrate.

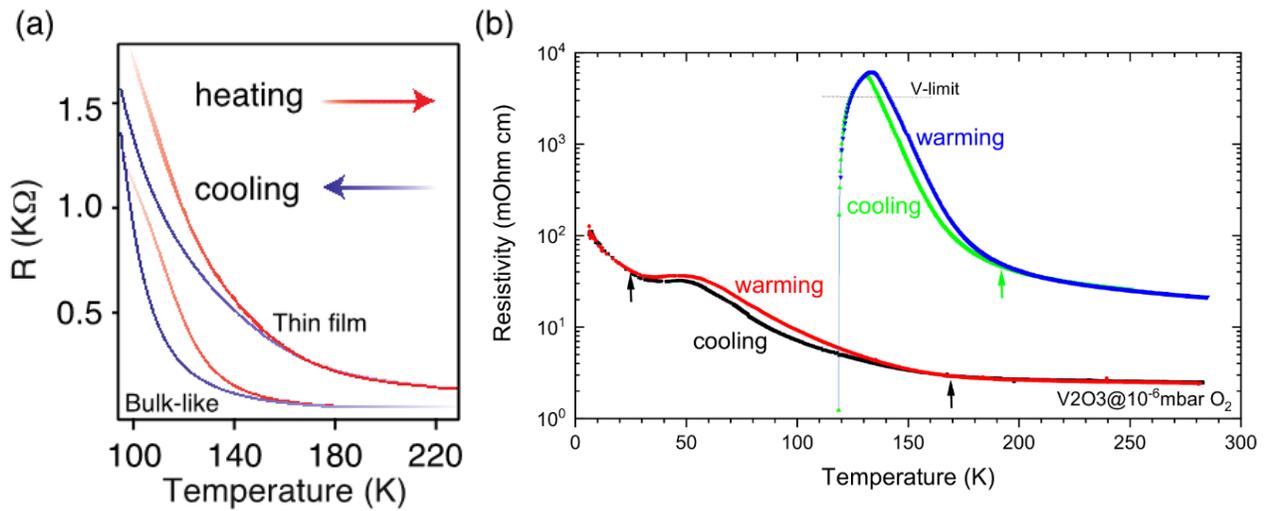

Figure S2. (a) Four-points probe resistivity measurements carried out for bulk-like films (40 nm) and thin films (~15 nm). For both, hysterical behaviour is observed within the minimum temperature allowed by our setup. With this we observed a strong up-rise, indicating that the samples are gradually approaching the insulating regime; (b) Resistivity measurements for samples grown with a mildly different oxygen pressure. In optimal doped sample (green/blue curves), closing of the hysteresis is just an artefact: The V-limit has been reached and the ability to measure the sample in a truly insulating phase is hampered. However, extra-oxygenated samples (black-red lines), though affected by a significant shift in the transition temperature at which the hysteresis opens, show however its full closure. The green arrow indicates the beginning of the hysteresis for the green-blue resistivity curve. Similarly, the black arrows indicate the beginning and the end of the hysteresis in the most conducting samples (black-red lines).

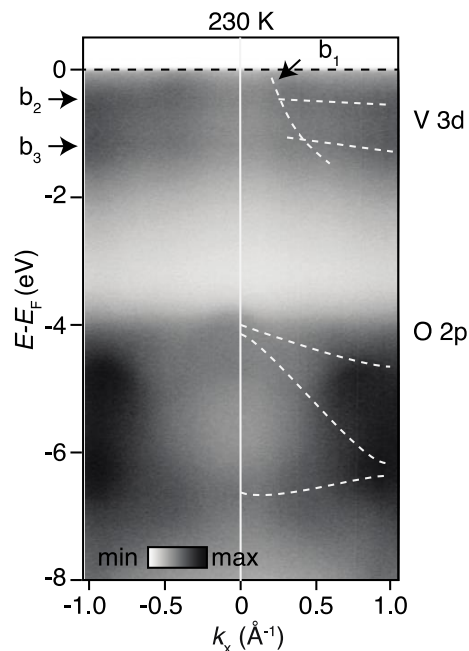

Figure S3. ARPES measurements along the Γ-K showing an overview of the bands that form the electronic structure of $V_2O_3$. The V3d states form the near-Fermi electronic structure, the O2p are located at lower values of binding energy with a maximum slightly above 4 eV. The dashed lines are guidelines for the eyes.

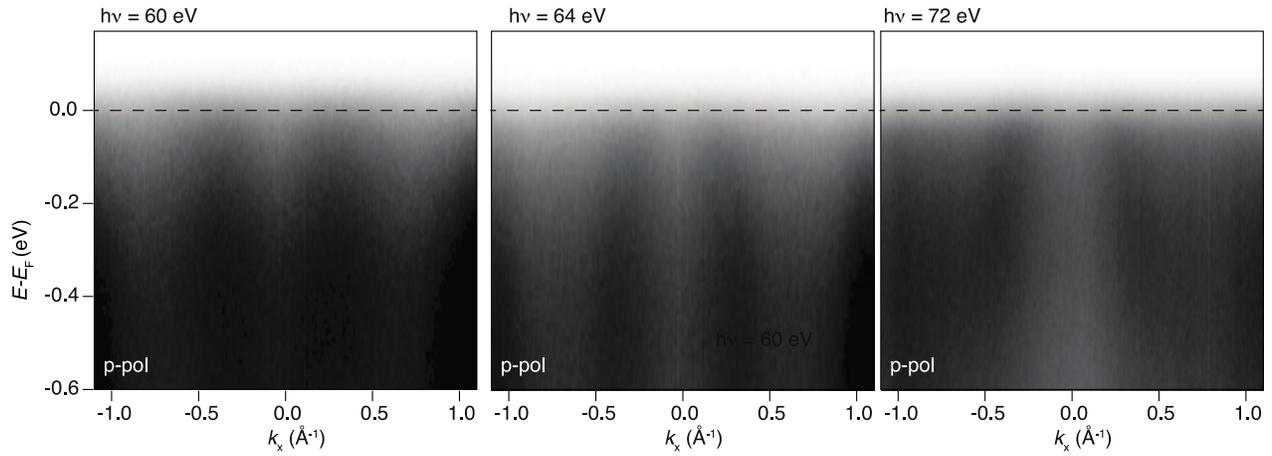

Figure S4. ARPES measurements along the Γ-K direction at different kz. The photon energies are collected such that the bands are shown along the Γ-Z direction and are indicated above the spectra. P-polarization is used to get information of both $e^\pi_g$ and $a_{1g}$ and to detect if any sign of an electron-like band is present. To note that the matrix elements show a very strong dependence as function of photon energy.

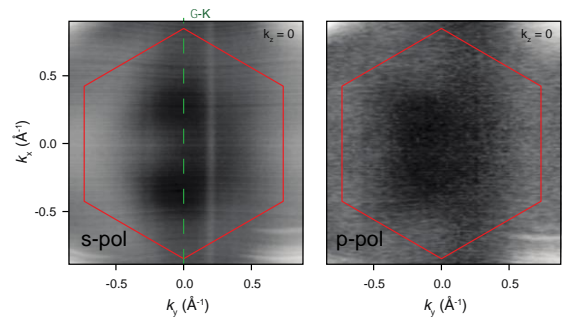

Figure S5. Fermi surface map collected at the plane $k_z=0$ (72 eV) showing the hexagonal section of the Brillouin zone (red) along with the direction of the ARPES data shown in the main text. The light polarisation used here was s-polarisation (left) and p-polarisation (right). The choice of this measurement high-symmetry direction is justified by the favourable matrix elements in our experimental setup for s-polarisation which strongly suppressed the intensity in the direction orthogonal to it. In this configuration instead, we were able to better disentangle the electronic structure of $V_2O_3$ and to better track the bands evolution throughout the data acquisitions. The p-polarisation in our geometry is in agreement with pervious ARPES [I. Lo Vecchio et al., Phys. Rev. Lett. 117, 166401 (2016)] and reverses the spectral way orthogonally compared to s-polarisation.



\bibliography{references}

\end{document}